# Optical Links for ATLAS Liquid Argon Calorimeter Front-end Electronics Readout

**Tiankuan Liu**[*]

*Southern Methodist University,
Dallas, Texas 75275, USA*
*E-mail*: liu@physics.smu.edu

ABSTRACT: We present the optical data links for the ATLAS liquid argon calorimeter. The current status of the vertical cavity surface emitting laser failures, the up-to-date results in searching for the failure cause, experiences gained in the searching process, possible backup plans for the optical transmitters and the lessons learned are also discussed.

---

[*] On behalf of the ATLAS Liquid Argon Calorimeter Group.

**Contents**

**1. Introduction**

Optical data links have been developed to transmit data from front-end electronics to the back end electronics for the ATLAS Liquid Argon (LAr) Calorimeter [1] since 1998 and have been functioning as designed since 2007. This system provides valuable experiences for developments of future optical links.

Since 2007, we have experienced vertical cavity surface emitting laser (VCSEL) failures. Despite great efforts so far we do not understand the cause of failures. Based on the optical spectra of VCSELs, a semi-empirical non-invasive measurement has been developed to assess future failures. At the same time, two backup designs have been proposed and are under development.

**2. Overview of optical links for the ATLAS LAr calorimeter**

The front-end electronics, including 1524 front-end boards (FEBs) hosted in 58 front-end crates (FECs), of the ATLAS liquid argon calorimeter is mounted on the detector. Each FEB contains 128 channels [2]. The back-end electronics is located in the counting room 150 meters away [3]. Between the front-end and the back-end electronics are 1524 optical links. Each optical link runs at 1.6 gigabit per second (Gbps) and transmit the data of 128 channels on one FEB.

Figure 1 is the block diagram of optical links for the ATLAS LAr calorimeter. The G-link transmitter (TX) and receiver (RX) are a serializer/deserializer chip set specially ordered from Agilent Technologies, Inc. SMUX, an application specific integrated circuit (ASIC) fabricated in DMILL process, is the interface between the G-Link TX and the data coming from analog-to-digital-converters (ADCs). The optical transmitter (OTX) and receiver (ORX) are custom assembled optical interface modules. All components inside an OTX and an ORX are commercial-off-the-shelf (COTS) devices. The laser diode used in the OTX is an oxide confined VCSEL mounted in a TO-can package. The front-end electronics, which operates in a radiation environment, must tolerate up to total ionizing dose (TID) of 3.41 kGy($SiO_2$), non-ionizing energy loss (NIEL) of $1.09 \times 10^{14}$ 1-MeV equivalent neutrons/cm$^2$, and total fluence of $1.53 \times 10^{13}$ (energy greater than 20 MeV) hadrons/cm$^2$ in 10-year life time, including safety factors [4].

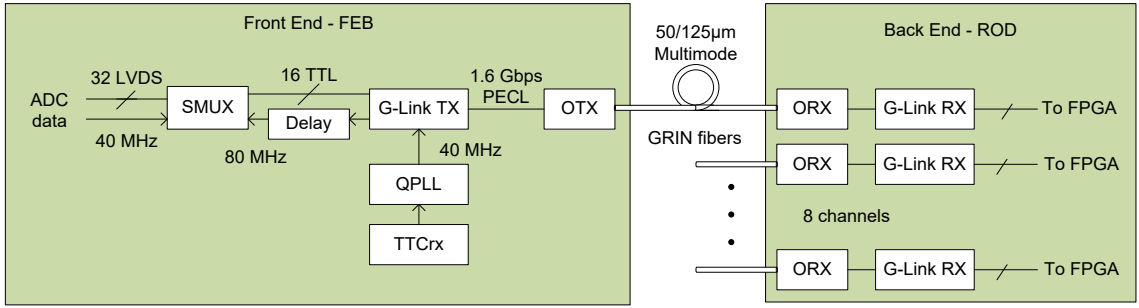

Figure 1. Block diagram of optical links for the ATLAS LAr Calorimeter

## 3. VCSEL failures

As mentioned in the previous section, in each optical link for the ATLAS LAr Calorimeter the data of 128 channels are transmitted. In other words, the failure of one optical link means the data loss of 128 channels. After the ATLAS LAr readout system started running at the beginning of 2007, few failures appeared until the first short run of the ATLAS detector in 2008. Since then, we have experienced optical link failures from the initial rate of one failure per week to the rate of one failure per month this year. The number of link failures and device power-on hours as functions of date are shown in Figure 2. The OTXs of the dead links were replaced in the 2008–2009 shut down and all failures were traced back to the VCSELs. Since May of 2009, we have had no access to the detector. By now there are twenty-four dead links. No dead link has light output, consistent with the symptoms of VCSEL failures. Therefore, we suspect that all optical link failures are due to VCSELs.

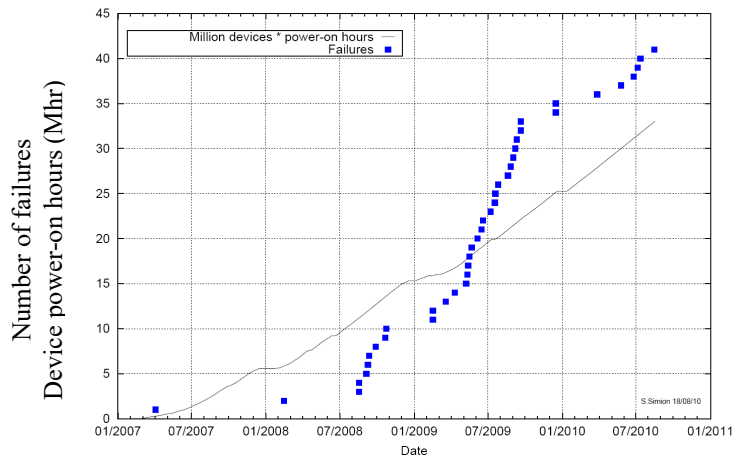

Figure 2. Link failures and device power-on hours as functions of date

A task force was established in December of 2008 to deal with the VCSEL issue. Two plans are proposed. Plan A is to replace the failed parts with similar devices if the cause is understood and removed, whereas Plan B is to replace all devices with new production. Since then, great efforts have been made to understand the failure cause. Several possible failure causes including humidity, magnetic field, electrostatic discharge (ESD), electrical overstress (EOS), and VCSEL fabrication defects, have been investigated [5]. So far we have not got to the root cause. It is worth mentioning that VCSEL failures exist not only in LAr, but also in other ATLAS sub-detectors and LHCb [6].

During the process searching for the failure cause, we have developed an empirical non-invasive measurement to assess future failures. Though we have no access to the front-end electronics during the ATLAS run, optical spectrum of the VCSEL in each optical link can still be measured in the counting room. Optical spectrum of a VCSEL becomes narrow after it is exposed to electrostatic discharge [7]. We verified that the same phenomenon occurs on the VCSELs used in the ATLAS LAr calorimeter. We chose randomly several spare OTXs and took the VCSELs out. Then we applied either forward or reverse ESD voltage pulses to each VCSEL. The optical spectrum of each VCSEL was measured before and after the ESD pulse application. Optical spectra of two VCSELs are shown in Figure 3. Above a certain threshold, the higher ESD voltage pulse is applied to a VCSEL, its optical spectrum becomes narrower. A VCSEL is more sensitive to reverse ESD than to forward ESD damage.

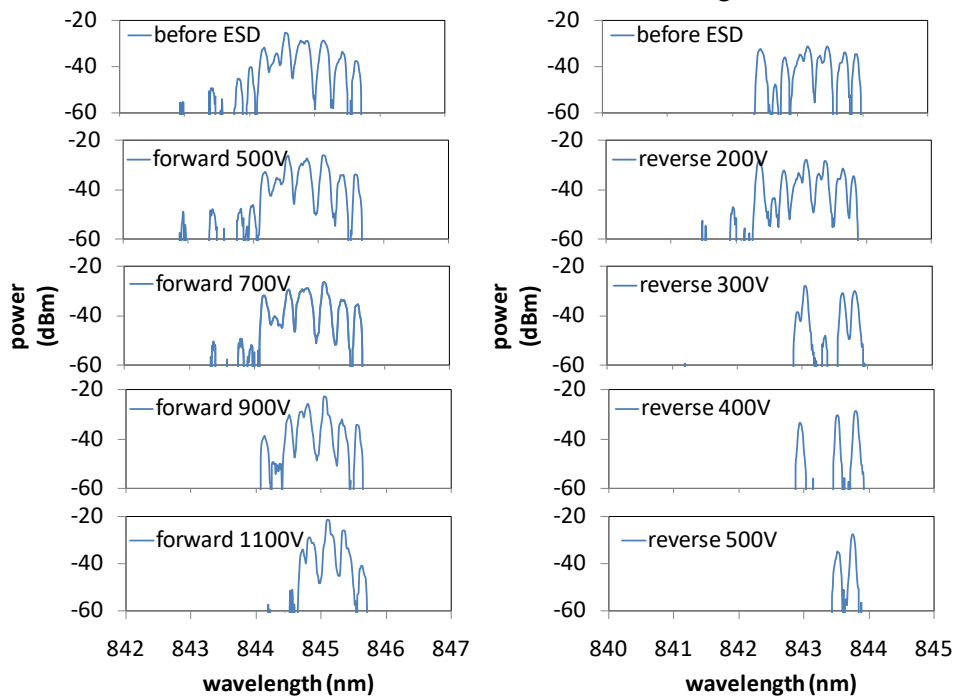

Figure 3. Optical spectra before and after an ESD damage

So far narrow optical spectrum method is effective at predicting which VCSEL is more likely to fail. A plot of spectral width versus OTX serial number is shown in Figure 4. The spectral width is defined as the narrowest window which contains 90% of the total power. All the points below -50 dBm or dBm/nm are ignored in the width calculation. Note that there is a serial number gap from 2200 to 4000. All failures are located in the narrow spectrum region except the one which is pointed to with a red arrow. We suspect this failure is not due to the VCSEL, but we cannot know for sure until we get the FEB out.

Optical spectra of VCSELs in the optical links are stable. We measured optical spectra four times. The spectral width differences between two measurements (February 2010 and July 2009) versus the spectral width in July 2009 are shown in Figure 5. Only one spectral width difference is larger than 0.5 nm because a peak at the far short wavelength end is included in the spectral width in one measurement and excluded in the other time.

While we look for the failure cause, we are developing two backup plans. The pictures of OTXs used at present and under development in the backup plans are shown in Figure 6. In both backup plans, the VCSELs used in the present OTXs will be replaced. The first backup plan (Plan B1) focuses on the compatibility. While the same laser driver chip is kept, a conventional biasing/coupling circuit is used to replace that used in the present OTX. No extra work needs to be done when we replace the present OTX with the new OTX in Plan B1. Plan B1 is still on the testing and qualification phase.

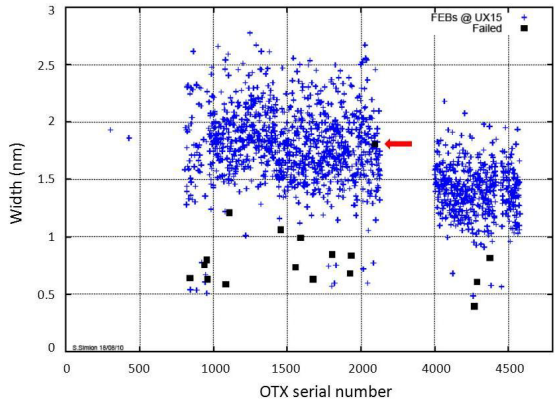
Figure 4. The OTX spectral width and serial number

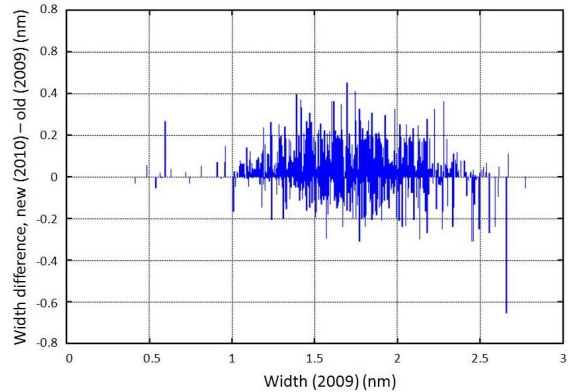
Figure 5. The OTX spectral stability

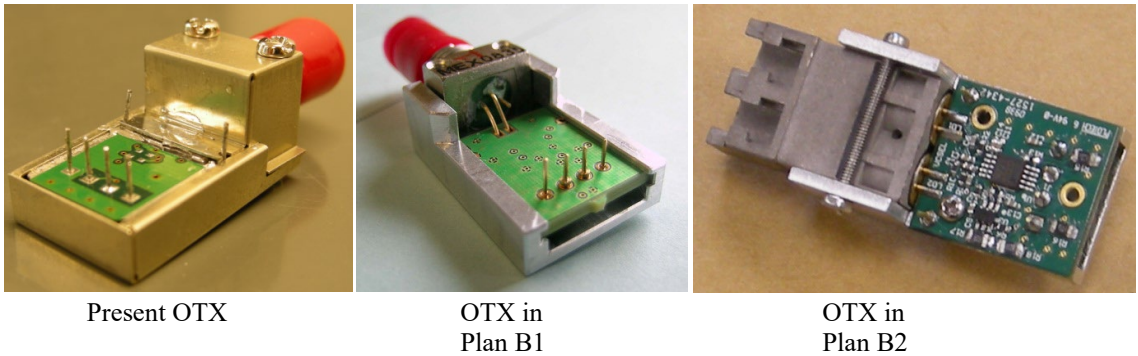

Present OTX　　　　　OTX in Plan B1　　　　　OTX in Plan B2

Figure 6. The pictures of present OTX, Plan B1 and Plan B2

The second backup plan (Plan B2) concentrates on redundancy. Each new OTX in Plan B2 contains two laser driver chips (only one driver chip can be seen in Figure 6), two VCSELs, and has two optical outputs. The laser drivers and the biasing/coupling circuits are the same as those in the present OTXs. The high speed input signals of each OTX drive both the laser drivers, but the signals are terminated at the far end only once. In order to keep the same module size, we use two smaller LC connectors in the new OTX to replace the single larger ST connector in the present OTX. In case one VCSEL dies, we only need to swap the two fibers in the control room. The replacement of new OTXs in Plan B2, compared to that in Plan B1, require extra work like modification on the front-panels of each FEB and the installation of extra fibers. However, the even bigger workloads in the replacement of new OTXs with either option is to get all FEBs out of the pit, dismount the cooling plates, and solder each OTX off an FEB.

In Plan B2 many tests including the radiation qualification have been done and the reliability test is still ongoing. 122 new OTXs have been assembled. Eye diagrams, average optical power, extinction ratio, rise/fall time, jitter, bit error rate (BER), and optical spectra of all channels have been measured except three modules with various problems. All measured parameters fall into quality control specifications of the present OTXs. Radiation qualification of new OTXs in Plan B2 has been done. Four new OTXs have been tested in 200 MeV proton beam. All OTXs continued to function throughout the test. For two modules, the power supply current decreased less than 1% with TID of 4.2 kGy($SiO_2$), NIEL of $6.72\times10^{12}$ 1-MeV equivalent neutrons/$cm^2$, and fluence of $7.06\times10^{12}$ protons/$cm^2$ and annealed less than 1% after 22 hours. For other two modules, the power supply currents changed less than 8% with TID of 133 kGy($SiO_2$), NIEL of $2.16\times10^{14}$ 1-MeV equivalent neutrons/$cm^2$, and fluence of $2.26\times10^{14}$ $cm^{-2}$ and annealed less than 1% after 15 hours. Cross section is estimated to be less than $1.8\times10^{-10}$ $cm^2$. Correspondingly, the estimated BER at ATLAS LAr is less than $5\times10^{-14}$, much less than the industrial standard $1\times10^{-12}$.

| Tuning frequency (GHz) | Upper | 4.98±0.02 |
|---|---|---|
| | Lower | 4.63±0.12 |
| At the central frequency | Power consumption (mW) | 111±8 |
| | Amplitude (pk-pk of differential, pk-pk, V) | 1.23±0.09 |
| | Rise time (20% - 80%, ps) | 44.9±2.4 |
| | Fall time (20% - 80%, ps) | 44.4±2.2 |
| | Random jitter (RMS, ps) | 1.3±0.3 |
| | Deterministic jitter (pk-pk, ps) | 7.5±1.1 |

Table 1. The radiation tolerant requirements and the measurements

## 4. Lessons learned in the development of ATLAS LAr optical links

We learned a lot of lessons in the development of ATLAS LAr optical links. System specifications should be studied in the beginning of R&D. We had not established any jitter requirement on the reference clock until an issue occurred. The importance of redundancy was not realized in the R&D. In fact, a dual channel redundancy scheme was developed in R&D, but declined in the final implementation due to the cost.

A pluggable module containing OTX or a mezzanine board containing an optical link will be desirable. At present all components of optical links were soldered on front-end board or readout driver boards. The replacement of an OTX may damage a whole FEB.

We cannot overemphasize the importance of reliability in a system without much access. In the production and quality assurance phase, we spent a lot of time to learn the OTX burn-in process. We performed a reliability test on OTX, but the total device hours were too small to catch any VCSEL failure.

## 5. Conclusion

Other than the VCSEL failure at a few percent level, the optical links for ATLAS LAr front-end electronics readout are functioning as designed and transmitting physics data. Although we have made many attempts to understand the cause of VCSEL failures, so far we failed to get to the root cause. However, narrow optical spectrum method is effective at predicting which OTX is likely to fail. Two backup plans are under development and will be production ready before the next LHC shutdown. At some point, we have to make a decision among replacing dead and narrow spectrum OTXs if spectrum indication is still effective or replacing all OTXs with new compatible design or redundant design. Among the lessons we have learned in the optical link development, the most profound lesson is that we cannot overemphasize the importance of reliability in a system without much access.

## Acknowledgments

This work is supported by the US ATLAS maintenance and operation fund. The authors would like to thank Joseph Hashem, Cotty Kerridge, Chonghan Liu, Sophia Wang, Yuan Zhang for measuring the present OTXs and the new OTXs with redundancy.